\documentclass[12pt]{article}
\newsavebox{\ns}
\newsavebox{\dbrane}
\newsavebox{\dbshort}

\usepackage{epsfig}
\usepackage{amssymb}

\def\appendix{{\newpage\section*{Appendix}}\let\appendix\section%
        {\setcounter{section}{0}
        \gdef\thesection{\Alph{section}}}\section}

\def\be{\begin{equation}}
\def\ee{\end{equation}}
\def\ba{\begin{eqnarray}}
\def\ea{\end{eqnarray}}

\def\Dslash{\,\,{\raise.15ex\hbox{/}\mkern-12mu D}}
\def\Dbarslash{\,\,{\raise.15ex\hbox{/}\mkern-12mu {\bar D}}}
\def\delslash{\,\,{\raise.15ex\hbox{/}\mkern-9mu \partial}}
\def\delbarslash{\,\,{\raise.15ex\hbox{/}\mkern-9mu {\bar\partial}}}
\def\pslash{\,\,{\raise.15ex\hbox{/}\mkern-9mu p}}
\def\calDslash{\,\,{\raise.15ex\hbox{/}\mkern-12mu {\cal D}}}

\newcommand\mr{\mathbb{R}}
\newcommand\mcp{\bf \mathbb{C}P}

\newcommand\Index{\mathrm{Index}}

\newcommand\Sign{\mathrm{Sign}}
\newcommand\cosec{\mathrm{cosec}}

\begin{document}

\begin{titlepage}

\begin{center}
\today
{\small\hfill hep-th/0209260}\\
{\small\hfill QMUL-PH-02-19}\\

\vskip 1.5 cm
{\large \bf Anomalous Couplings in M-Theory and String Theory}

\vskip 1 cm
{James Sparks}\\
\vskip 1cm

{\sl Department of Physics \\ 
Queen Mary, University of London \\
Mile End Road, London E1 4NS, UK \\
{\tt J.Sparks@qmul.ac.uk}\\}

\end{center}

\vskip 0.5 cm
\begin{abstract}

We explore how various anomaly-cancelling terms in M-theory and string theory transform non-trivially into each other under duality. Specifically, we study the phenomenon in which bulk terms in M-theory get mapped to brane worldvolume terms in string theory. The key mathematical ingredient is $G$-index theory.

\end{abstract}

\end{titlepage}

\pagestyle{plain}
\setcounter{page}{1}
\newcounter{bean}
\baselineskip18pt

%%%%%%%%%%%%%%%%%%%%%%%%%%%%%%%%%%%%%%%%%%%%%%%%%%%%%%%%%%%%%%%%%%%%%

\section{Introduction and Summary}

In this note we investigate how various anomalous couplings in
M-theory and string theory are related to each other under
duality. In particular, we focus on the phenomenon in which bulk terms in M-theory get mapped to brane worldvolume couplings in string theory. 

In order to explain this, consider some configuration of D6-branes in type IIA string theory. The D6-brane is a Kaluza-Klein monopole and so naively this configuration will lift to some geometric background in M-theory. In the effective worldvolume theory of the D6-branes there are Wess-Zumino terms which couple the brane to the bulk Ramond-Ramond fields. These couplings are related to anomaly cancellation for intersecting branes, and, in certain circumstances, can lead to induced topological charges. A natural question to ask is, therefore, what happens to these couplings when we lift to M-theory?

Alternatively, we may pose the problem the other way around. As an illustration, consider the gravitational Chern-Simons term in M-theory

\be
\int C\wedge I_8(\mathcal{R})\label{CS}\ee

Here $C$ denotes the three-form potential and $I_8(\mathcal{R})$ is a certain eight-form constructed out of the Riemann tensor $\mathcal{R}$. The term (\ref{CS}) participates in an anomaly inflow mechanism which cancels an
anomaly in the M5-brane worldvolume theory \cite{duff, wittenm5}. Reducing to type IIA string theory on a circle, (\ref{CS}) trivially becomes \cite{duff} the one-loop interaction found in \cite{vafawitten} on replacing $C$ with the NS-NS two-form $B$ which couples to the fundamental string. This interaction was originally calculated by compactifying the IIA string on an eight-dimensional manifold $X$ and then computing the one-point function of the $B$-field \cite{vafawitten}.

However, we shall argue here that there is a non-trivial reduction of the anomalous term (\ref{CS}) to type IIA which is not a Kaluza-Klein reduction in the usual sense. To explain this, recall \cite{Ibrane} that on the worldvolume $W$ of a D6-brane one has a coupling

%the effective worldvolume theory of a D-brane contains a Wess-Zumino term which describes the coupling of the D-brane to the bulk Ramond-Ramond (RR) fields

%\be
%I_{\mathrm{WZ}} = \int_W {C_*} \wedge \mathrm{ch}(E)
%\wedge \sqrt{{\hat A (TW) \over \hat A (NW)}}
%\label{anomwz}
%\ee

%Here $C_*$ denotes the formal sum of RR potentials, $\mathrm{ch}(E)$ is the Chern character of the gauge bundle $E$, $TW$ denotes the tangent bundle of the brane worldvolume $W$, $NW$ denotes the
%normal bundle of $W$ in spacetime, and $\hat {A}$ is the Dirac genus. The coupling (\ref{anomwz}) is required to cancel anomalies that arise when branes intersect, again via an anomaly inflow mechanism \cite{Ibrane}, and may also be deduced via a perturbative string calculation \cite{stringpert}. In particular, on a D6-brane worldvolume $W$ one has a coupling

\be
\int_W {C_3} \wedge \frac{1}{48}\left( -p_1(TW) + p_1(NW)\right)\label{gravanom}
\ee

where $C_3$ denotes the (pull-back of the) Ramond-Ramond (RR) three-form potential, $TW$ denotes the tangent bundle of the brane worldvolume $W$, $NW$ denotes the normal bundle of $W$ in spacetime, and $p_1$ is the first Pontryagin class. We will argue that the terms (\ref{CS}) and (\ref{gravanom}) are in fact dual to each other. This is clearly not a standard Kaluza-Klein reduction, in the sense that the $C I_8$ coupling in M-theory reduces to $B I_8$ in string theory, for example. 

One may argue that (\ref{CS}), (\ref{gravanom}) are related by a simple extension of some arguments that may be found in \cite{vafawitten, BVS}. Start in type IIA and consider the interaction $B I_8$. If one ``compactifies'' the IIA string to $M_6$ on the ALE singularity $A_{N-1}$, one finds that this one-loop term gives rise to a quantum correction \cite{BVS}

\be
-\frac{N}{48}\int_{M_6} B \wedge p_1\ee

One may now perform a ``9-11 flip'', trivially lifting to M-theory on a circle ${\bf S}^1$, and coming down to type IIA on an ${\bf S}^1$ contained in the ALE space. One ends up with $N$ D6-branes with worldvolume $M_6\times {\bf S}^1$ and a brane coupling

\be
-\frac{N}{48}\int_{M_6\times {\bf S}^1} C_3 \wedge p_1\ee

This is precisely the gravitational Wess-Zumino coupling (\ref{gravanom}). Unfortunately, there is no contribution from the normal bundle in this case, since the latter is trivial. Moreover, the computation is entirely local - that is, we have only considered a neighbourhood of the D6-brane.

In this note we prove equality of the couplings (\ref{CS}) and (\ref{gravanom}) in a more general setting. As in \cite{vafawitten}, we consider the case of compactification on an arbitrary eight-dimensional spin manifold $X$. Equality of the gravitational terms (\ref{CS}), (\ref{gravanom}) then follows using $G$-index theory on $X$.  A $G$-index theorem is just like a usual index theorem for an elliptic operator, but with an additional group action by $G$, and may be regarded as a
generalisation of the Lefschetz fixed point formula. We will take $G\cong U(1)$, which generates motions around the M-theory circle fibres. The beauty of this approach is that we prove equality of the gravitational terms in general (in the context of compactification on $X$), corresponding to an arbitrary configuration of D6-branes in type IIA. 

Similarly, we argue using anomaly cancellation that the Chern-Simons interaction of eleven-dimensional supergravity maps to the gauge Wess-Zumino terms for a single D6-brane. Thus, summarising (in conventions described in the next section), we therefore have what might be described as ``non-standard'' Kaluza-Klein reductions:

\be
\int_{M_{11}} C\wedge I_8(\mathcal{R}) \sim \int_W C_3 \wedge \frac{1}{48}\left( -p_1(TW) + p_1(NW)\right)\ee

\be
\frac{1}{24\pi^2}\int_{M_{11}} C\wedge G\wedge G \sim \int_W C_3 \wedge \mathrm{ch}(E)\ee

from M-theory to type IIA string theory with D6-branes, where $\mathrm{ch}(E)$ denotes the Chern character of the gauge bundle $E$. Moreover, by applying $T$-duality one recovers the gravitational and gauge couplings for general D$p$-branes in type II string theory.

The plan of the paper is as follows. In section 2 we consider compactification of M-theory on $X$, focusing in particular on the membrane tadpole, and the dual interpretation in type IIA string theory. In section 3 we work out the relevant $G$-index theorems and use the results to prove the relations above.

%%%%%%%%%%%%%%%%%%%%%%%%%%%%%%%%%%%%%%%%%%%%%%%%%%%%%%%%%%%%%%%%%%%%%%%%

\section{Effective Membrane Charges}

Our starting point is the low-energy effective action for the bosonic
fields of M-theory. This takes the form

\be
S = \frac{1}{2\kappa^2_{11}}\int \left(d^{11}x \sqrt{-g}R-\frac{1}{2}G\wedge *G -
\frac{1}{6}C\wedge G\wedge G\right)-\int C\wedge I_8(\mathcal{R})\label{maction}\ee

where $\kappa^2_{11}$ is related to the eleven-dimensional Planck
length by $\kappa^2_{11}\sim l_P^9$. For a careful
treatment of the numerical factors see, for example,
\cite{alwis}. We shall set the M2-brane tension $T_2$ to unity. It then
follows that the M5-brane tension is given by $T_5=1/2\pi$ and
$\kappa^2_{11}=2\pi^2$. 

The last term in (\ref{maction}) is a gravitational Chern-Simons
correction to the eleven-dimensional supergravity action. Explicitly, we have

\be
\int C\wedge I_8(\mathcal{R}) = \frac{1}{192}\int C\wedge \left(p_1^2-4p_2\right)\ee

The Pontryagin forms are given in terms of the curvature two-form $\mathcal{R}$ by

\be
p_1 = -\frac{1}{8\pi^2}\mathrm{tr}\mathcal{R}^2, \quad p_2 =
-\frac{1}{64\pi^4}\mathrm{tr}\mathcal{R}^4 + \frac{1}{128\pi^4}(\mathrm{tr}\mathcal{R}^2)^2\ee

In order to ``see'' the gravitational Chern-Simons term, as in \cite{vafawitten} we consider
compactification of M-theory on a closed eight-dimensional spin manifold
$X$. As pointed out in \cite{GS}, eight dimensions is particularly interesting as far as anomalies are concerned, which is partly why we focus attention on this case. Thus the M-theory background is of the form $M_3\times X$ where
$M_3$ is a three-manifold. One might typically take
$M_3=\mr^{2,1}$, for example. Variation of the $C$-field gives
rise to the equation of motion

\be
\frac{1}{4\pi^2}d*G + \sum_{i=1}^{N_{M2}}\delta(x_i) +\frac{1}{192}\left(p_1^2(X)-4p_2(X)\right)+
\frac{1}{2}\left(\frac{G}{2\pi}\right)^2=0\label{Ceom}\ee

where we have included $N_{M2}$ space-filling
M2-branes in the vacuum. The $i^{\mathrm{th}}$ M2-brane is located at the point $x_i\in X$, and $\delta(x_i)$ denotes the Poincar\'e dual to the worldvolume\footnote{For an anti-M2-brane one has a minus sign.}.

Integrating (\ref{Ceom})
over $X$ gives rise to the anomaly cancellation condition

\be
N_{M2}+
\frac{1}{192}\int_X \left(p_1^2-4p_2\right)+\frac{1}{2}\int_X \left(\frac{G}{2\pi}\right)^2 =0\label{anom}\ee

The fact that, given an arbitrary spin manifold $X$ together with a correctly quantised $G$-flux, one can always solve (\ref{anom}) with $N_{M2}$ integer is actually rather non-trivial \cite{wittenflux}.

Suppose now that there exists a reduction of M-theory on $X$ to type
IIA string theory on $Y$, with D6-branes in the vacuum. This means that
$X$ admits a semi-free circle action\footnote{That is, freely acting outside the set of points fixed under the entire group action.} with a codimension four fixed
point set $B$. Notice that $B$ need not be connected. $X$ is a circle bundle over $Y$, with the circle
fibres degenerating over the locus $B$. This circle bundle is usually called the ``M-theory circle bundle''. 

More precisely, we may construct the type IIA manifold $Y$ as follows. We start by deleting from $X$ a small tubular neighbourhood surrounding $B$. The boundary of that tubular neighbourhood is a seven-manifold $P$ which is a three-sphere bundle over $B$. When we reduce to type IIA by quotienting out the M-theory circle, the boundary $P$ maps to a boundary six-manifold $S=P/U(1)$ which is a two-sphere bundle over $B$. We complete the IIA spacetime $Y$ by
filling in each two-sphere fibre with a three-disc, thus obtaining a closed seven-manifold.

M-theory on $X$ is then dual to type IIA string theory on $Y$ with
a space-filling D6-brane wrapped over the codimension three submanifold
$B\subset Y$, where $B$ is embedded as the zero-section of the disc bundle described in the last paragraph. Thus if ${\bf S}^2$ denotes a small two-sphere linking a
connected component of $B$ in $Y$ we have

\be
\int_{{\bf S}^2} \frac{G_2}{2\pi} = 1\label{oned6}\ee

where $G_2$ denotes the RR two-form field strength. After reduction to type IIA theory the effective membrane
charges become the effective charge of space-filling D2-branes.
Thus there must be a type IIA interpretation of the anomaly formula (\ref{anom}).

We assume throughout that the NS-NS $H$-field is cohomologically
trivial. This means that the integral of the M-theory $G$-flux over
the M-theory circle is zero in $H^3(Y)$. There is therefore no contribution to the D2-brane
charge from the bulk.
However, in type IIA theory we also have a space-filling
D6-brane wrapped over the four-cycle $B\subset Y$.
Due to the non-trivial embedding of the D6-brane worldvolume
in spacetime, the Ramond-Ramond fields in the bulk couple to various terms on the D6-brane worldvolume \cite{Ibrane}

\be
I_{\mathrm{WZ}} = \int_W {C_*} \wedge \mathrm{ch}(E)
\wedge \sqrt{{\hat A (TW) \over \hat A (NW)}}
\label{anomwz}
\ee

where $\hat A$ denotes the Dirac genus. Comparing the $C_3$ coupling in (\ref{anomwz}) with the formula (\ref{anom}), we see that the type
IIA analogue of the latter is

\be
N_{D2} + \int_B  \sqrt{{\hat A (TB) / \hat A (NB)}} +
\frac{1}{2}\int_B \left(\frac{\mathcal{F}}{2\pi}\right)^2\label{2anom}=0
\ee

where $N_{D2}$ is the number of space-filling D2-branes, and the $U(1)$ gauge field strength on the D6-brane is denoted $\mathcal{F}$. $N_{D2}$ is
naturally identified with $N_{M2}$ in M-theory. The aim of this paper
is to justify that the geometric (and flux) terms are also naturally
identified. That is

\be
\frac{1}{192}\int_X \left(p_1^2(X)-4p_2(X)\right) = \frac{1}{48}\int_B \left(-p_1(TB)+p_1(NB)\right)\label{geom}\ee

and

\be
\frac{1}{2}\int_X \left(\frac{G}{2\pi}\right)^2 = \frac{1}{2} \int_B
\left(\frac{\mathcal{F}}{2\pi}\right)^2\label{flux}\ee

The relation (\ref{geom}) between the Pontryagin numbers of $X$ and
the Pontryagin numbers of $B$ will be proved in the next section using
a combination of $G$-index theorems. The relation (\ref{flux}) between the $G$-flux in M-theory and the
$\mathcal{F}$-flux on the D6-branes then follows from anomaly cancellation in M-theory and string theory. That is, after identifying $N_{M2}$ with $N_{D2}$ and having proven (\ref{geom}), the relation (\ref{flux}) then follows from anomaly cancellation. In fact, one can show quite generally that the gauge field on a D6-brane is determined in terms of the RR four-form by the formula

\be
\frac{\mathcal{F}}{2\pi} = \int_{{\bf S}^2} \frac{G_4}{2\pi}\ee

where ${\bf S}^2$ is as in (\ref{oned6}), and that the Freed-Witten quantisation
law for $\mathcal{F}$ then follows from subtle factors in
the quantisation law for $G_4$. Details may be found in \cite{thesis}.

%%%%%%%%%%%%%%%%%%%%%%%%%%%%%%%%%%%%%%%%%%%%%%%%%%%%%%%%%%%%%%%%%%%%%

\section{Gravitational Chern-Simons Terms and $G$-Index Theorems}

Our aim in this section is to prove the formula (\ref{geom}) relating the gravitational Chern-Simons term in M-theory with the
gravitational Wess-Zumino terms on a D6-brane in type IIA string
theory. As we shall see, this is a consequence of a combination of
$G$-index theorems. 

The relation between index theory and anomalies is of course well-known, but $G$-index theory is perhaps less familiar to physicists\footnote{There are some notable exceptions in the literature. For example, a $\mathbb{Z}_2$-index theorem was used in \cite{scrucca} to compute the anomaly for a self-dual tensor.}. We shall show here that anomalous terms in M-theory and anomalous couplings in string theory may be related via $G$-index
theorems, where $G\cong U(1)$ acts by rotating around the M-theory circle direction. 

It is perhaps best to start with a simple example of what we mean by a
$G$-index theorem. We take $G\cong U(1)$, since this will be the case
of interest. Suppose then that we have a circle action on an oriented
even-dimensional manifold $X$, with
fixed point set $F$. The latter need not be connected. Each connected
component will have even codimension, although the dimensions of different components may differ. Then the Euler number of $X$ may be written as
the sum of the Euler numbers of each connected component $F_i$ of $F$

\be
\chi(X) = \sum_i \chi(F_i)\ee

where the index $i$ runs over each connected component. This is an
example of a $G$-index theorem. The elliptic operator is just the
exterior derivative. As a simple example of this particular theorem, take $X={\bf S}^2$
with circle action $U(1)\subset SO(3)$. This rotates the two-sphere,
leaving fixed the north and south poles. The Euler number of a point
is 1, and so we get the well-known result that

\be
\chi({{\bf S}^2}) = \chi(\mathrm{north\ pole})+ \chi(\mathrm{south\
pole}) = 1 + 1 = 2\ee

Since the group $U(1)$ is connected, the general $G$-index theorem of
\cite{AS} may be used to relate the index of an elliptic operator to various
characteristic classes evaluated on the fixed point set. A codimension four fixed point set $B$ then gets interpreted as a D6-brane worldvolume in type IIA string theory. We now show how this works in detail for two elliptic operators on the closed
spin eight-manifold $X$. 

%%%%%%%%%%%%%%%%%%%%%%%%%%%%%%%%%%%%%%%%%%%%%%%%%%%%%%%%%%%%%%%%%%%%%%5

\subsection{The $G$-Signature Theorem}

Recall that the signature of a compact oriented eight-manifold $X$ is defined to be the
signature of the quadratic form defined on $H^4(X;\mathbb{Z})$ given by the cup
product

\be
H^4(X;\mathbb{Z})\ni u \mapsto \int_X u\cup u\ee

Thus it is the number of (linearly independent) self-dual
harmonic four-forms minus the number of (linearly independent) anti-self-dual harmonic four-forms. We denote
the signature as $\Sign(X)$. This may be expressed as the index of an elliptic operator on $X$
\cite{AS}. There is also a $G$-index theorem for this operator
where we have in addition an action by the group $G$. The form of this $G$-index
theorem was worked out in the original paper \cite{AS}. In order to
state it, we will first of all need to establish some basic
notations. 

The action of $U(1)$ on $NB$, the normal bundle of the fixed point set
$B$ in $X$, induces a natural complex structure on $NB$. Indeed, $NB$
splits into a sum of two complex line bundles, and we take the $U(1)$
action on each to be multiplication by $e^{i\theta}$, where $\theta\in [0,2\pi]$ is the $U(1)$ group parameter. Thus $NB$,
with its $U(1)$ action, is naturally a rank
two complex vector bundle over $B$. We shall denote this bundle as
$V=NB$. Thus $V$ has Chern classes. These will be denoted by
$c_1=c_1(V)\in H^2(B;\mathbb{Z})$ and
$c_2=c_2(V)\in H^4(B;\mathbb{Z})$. 

Then the $G$-signature theorem of \cite{AS} in this case reduces to

\be
\Sign(X) = -\cot^2(\theta/2) \int_B L(TB) \wedge \mathcal{M}^{\theta}\ee

Here $L$ is the usual Hirzebruch $L$-polynomial. It may be written
in terms of Pontryagin classes as\footnote{This is the usual
definition. The definition in \cite{AS} differs by factors
of 2. Our definition of $\mathcal{M}^{\theta}$ also therefore differs
by factors of 2.}

\be
L = 1 + \frac{1}{3}p_1 + \frac{1}{45}\left(-p_1^2 + 7p_2\right) + \ldots\ee

The stable characteristic class
$\mathcal{M}^{\theta}=\mathcal{M}^{\theta}(c_k)$ is a polynomial
in the Chern classes of $V$. The coefficients are functions of
$\theta$. Specifically, we calculate\footnote{The $c_1$ coefficient was computed in \cite{AS}.}

\be
\mathcal{M}^{\theta} = 1 + 2i\cosec(\theta) c_1 - \cosec^2(\theta/2) (c_1^2-2c_2) - 4\cosec^2(\theta) c_2+\ldots\ee

The $G$-signature theorem thus gives

\begin{eqnarray}
\Sign(X) & = & -\cot^2(\theta/2) \cdot \Sign(B) + \cosec^2(\theta/2)
\cot^2(\theta/2) \int_B (c_1^2-2c_2) \nonumber \\
& & + \cosec^4(\theta/2) \int_B c_2\label{Gsign}\end{eqnarray}

where we have used the ordinary signature theorem to write

\be
\Sign(B) = \frac{1}{3}\int_B p_1(TB)\ee

The left-hand side of (\ref{Gsign}) is clearly independent of the group parameter
$\theta$. Thus the right-hand side should also be independent of
$\theta$. By performing a Laurent expansion of the right-hand side of
(\ref{Gsign}) around $\theta=0$, we find that setting the coefficients
of the $\theta^{-4}$ and $\theta^{-2}$ terms to zero gives the
constraints

\be
\int_B c_1^2 = \int_B c_2\label{con1}\ee

and

\be
\Sign(B) = \int_B c_2\label{con2}\ee

respectively, where a sum over each connected component of $B$ is understood in
these formulae. The constant term in the Laurent expansion then yields
the $G$-signature theorem:

\be
\Sign(X) = \Sign(B)\ee

In terms of Pontryagin forms, this reads

\be
\frac{1}{45} \int_X \left(-p_1^2(X)+7p_2(X)\right) = \frac{1}{3}\int_B p_1(TB)\ee

Unfortunately, this is not the correct linear combination of
Pontryagin numbers required to prove the relation between the
anomalous terms (\ref{geom}). We shall therefore
need another $G$-index theorem. It is straightforward to check that 

\be
\frac{1}{192}\int_X \left(p_1^2(X)-4p_2(X)\right) = 2\cdot \Index(\mathrm{Dirac}(X)) - \frac{1}{8}\Sign(X)\ee

in general. Thus we need to study the $G$-index theorem for the Dirac operator on $X$.

\subsection{The $G$-Dirac Theorem}

We take the Dirac operator on the spin
manifold $X$. The
details of the $G$-index theorem in this case were not worked out explicitly in \cite{AS}. Rather than present the
details (which involve fairly standard calculations), I shall simply state the result. With
appropriate conventions, the $G$-index theorem reduces to

\be
\Index(\mathrm{Dirac}(X)) = -\frac{1}{4}\cosec^2(\theta/2) \int_B
\hat{A}(TB)\wedge \mathcal{C}^{\theta}\ee

Here $\hat{A}$ is the Dirac genus, given by

\be
\hat A = 1 - {p_1 \over 24} + {7p_1^2 - 4 p_2 \over 5760} + \ldots\label{genus}
\ee

and the stable characteristic class
$\mathcal{C}^{\theta}=\mathcal{C}^{\theta}(c_k)$ is again a polynomial
in the Chern classes of $V$. Specifically, we find that

\be
\mathcal{C}^{\theta} = 1 + \frac{1}{2}i\cot (\theta/2) c_1 +
\frac{1}{4}\cosec^2 (\theta/2)(c_2-c_1^2)+ \frac{1}{8}c_1^2+\ldots\ee

Thus the $G$-Dirac theorem reads

\begin{eqnarray}
\Index(\mathrm{Dirac}(X)) & = & \frac{1}{96}\cosec^2(\theta/2) \int_B p_1(TB)
- \frac{1}{32}\cosec^2(\theta/2) \int_B c_1^2 \nonumber \\
& & - \frac{1}{16}
\cosec^4(\theta/2)\int_B (c_2-c_1^2)\label{Gdirac}\end{eqnarray}

Again, we may perform a Laurent expansion of the right-hand side of
(\ref{Gdirac}) around $\theta=0$. Setting the coefficients
of the $\theta^{-4}$ and $\theta^{-2}$ terms to zero gives the
same constraints (\ref{con1}, \ref{con2}) as the $G$-signature theorem, with the final
result that the index of the Dirac operator is zero:

\be
\Index(\mathrm{Dirac}(X)) = 0\ee

%Combining this result with the usual index theorem for the Dirac
%operator therefore gives

%\be
%\Index(\mathrm{Dirac}(X)) = \frac{1}{5760}\int_X \left(7p_1^2(X)-4p_2(X)\right) = 0\ee

Putting everything together, we get

\be
\frac{1}{192}\int_X \left(p_1^2(X)-4p_2(X)\right) = -\frac{1}{8}
\Sign(X) = -\frac{1}{8} \Sign(B) = - \frac{1}{24} \int_B p_1(TB)\ee

Thus the result (\ref{geom}) follows if we can show that

\be
\int_B p_1(TB) = -\int_B p_1(NB)\ee

Again, a sum over each connected component of $B$ is understood in this formula. Thus it remains to prove the last equation, or, equivalently, that

\be
\int_B p_1(TY) = 0\label{charge}\ee

Equivalence follows since $TY\mid_B = TB\oplus NB$ and therefore $p_1(TY)\mid_B = p_1(TB) + p_1(NB)$. Assuming for the moment the result (\ref{charge}), together with anomaly cancellation in M-theory and string theory, we find that the $C$-field tadpole is given by

\be
\frac{1}{192}\int_X \left(p_1^2-4p_2\right)+\frac{1}{2}\int_X \left(\frac{G}{2\pi}\right)^2 =  \int_B  \mathrm{ch}{\mathcal{F}} \wedge \hat {A} (TB) = \mathrm{Index}(\mathrm{Dirac}(B))\ee

the index of the natural Dirac operator on $B$ which couples to the $U(1)$ gauge field\footnote{More accurately, the ``$U(1)$ gauge field'' is actually a $\mathrm{spin}^c$ connection.}. This is clearly an integer. One may therefore cancel the tadpole for $C$ by including $|\mathrm{Index}(\mathrm{Dirac}(B))|$ space-filling M2/D2 branes or anti-branes (depending on the sign of the index) in the vacuum.

There are at least two ways of proving (\ref{charge}), which essentially amounts to a statement of charge conservation. The first is an ``M-theory'' proof, where we use the constraints (\ref{con1}, \ref{con2}) of the $G$-index theorems. The second is a ``string theory'' proof, which uses only (co)-bordism theory.

\subsubsection*{M-theory proof}

The first step is to write the Pontryagin class $p_1(NB)$ in string
theory in terms of Chern classes in M-theory. That is, we wish to
relate a characteristic class of the normal bundle of $B$ in $Y$ to
characteristic classes of the normal bundle of $B$ in $X$, which
recall we denoted by $V$. 

We may do this as follows. The structure
group of the rank two complex vector bundle $V$ is $U(2)$. Choose the
standard maximal torus $T=({\bf S}^1)^2$ for $U(2)$ with basic characters $x_1,x_2$. Then
reducing from M-theory to type IIA corresponds to taking the
projectivisation of this structure group. Thus the structure group of
the normal bundle of $B$ in $Y$ is $SO(3)\cong U(2)/U(1)$ where the
$U(1)$ denotes the central $U(1)$ subgroup. The
determinant line bundle of $V$ has first Chern class
$c_1(V)=x_1+x_2\in H^2(B;\mathbb{Z})$. It follows that the basic character for the maximal torus
$T={\bf S}^1$ for $SO(3)$ may be taken to be the anti-diagonal
combination $x_1-x_2$. 

Recall now that the first Pontryagin class of a real vector bundle $E$ may be
defined in terms of the basic characters $x_i$ of the standard maximal torus
of $SO(n)$ as

\be
p_1(E) = x_1^2+\ldots+ x_m^2\ee

where $m=\lfloor\frac{n}{2}\rfloor$. Thus

\be
p_1(V_{\mathbb{R}}) = x_1^2 + x_2^2\ee

and 

\be
p_1(NB) = (x_1-x_2)^2\ee

It follows that

\be
p_1(NB) = 2p_1(V_{\mathbb{R}}) - c_1^2(V)\ee

Now, we also have

\be
p_1(V_{\mathbb{R}}) = -c_2(V_{\mathbb{R}} \otimes_{\mathbb{R}}
\mathbb{C}) = -c_2(V\oplus\bar{V}) = -2c_2(V)+c_1^2(V)\ee

and so

\be
p_1(NB) = 2(-2c_2(V)+c_1^2(V))-c_1^2(V) = -4c_2+c_1^2\ee

Finally, we may use the constraints (\ref{con1}, \ref{con2}) from the
$G$-index theorems to relate $c_2$, $c_1^2$ and $\Sign(B)$. We obtain

\be
\int_B p_1(NB) = \int_B -4c_2 + c_1^2 = -3\int_B c_2 = -3\cdot\Sign(B) = -\int_B p_1(TB)\ee

thus proving (\ref{charge}).

\subsubsection*{String theory proof and bordism theory}

We have now completed our task. However, the proof just given relied on the mathematical construction relating string theory with M-theory, together with $G$-index theory. There is actually a very nice ``elementary'' interpretation of (\ref{charge}) in terms of bordism theory. Specifically, (\ref{charge}) corresponds to the fact that a certain bordism invariant must vanish. We finish with a brief explanation.

Recall that the spin six-manifold $S$ bounds a small tubular neighbourhood around the D6-brane $B$ \footnote{We suppress the three-dimensional spacetime $M_3$ in the argument.}, so that $S$ is a two-sphere bundle over $B$. Over any fibre of this bundle, we have

\be
\int_{{\bf S}^2} \frac{G_2}{2\pi} = 1\label{oned6again}\ee

which states that there is a single D6-brane wrapped on $B$. Now, the RR two-form $a\equiv [G_2/2\pi]$ may be viewed as an element of $H^2(S;\mathbb{Z})$ that extends over $Y\setminus B$. This statement simply says that the only D6-branes present are those wrapped on $B$ (which is implicit in our notation). It follows that the pair $(S,a)$ is zero as an element of the spin bordism group $\Omega_6^{\mathrm{spin}}(K(\mathbb{Z},2))$. This group is non-zero, and to say that $(S,a)=0\in \Omega_6^{\mathrm{spin}}(K(\mathbb{Z},2))$ leads to the constraint (\ref{charge}).

To explain this in a little more detail, recall that the spin bordism group $\Omega_n^{\mathrm{spin}}(Z)$ of the space $Z$ may be defined by taking pairs $(M_n,f)$, where $f:M_n \rightarrow Z$ is a continuous map from the spin $n$-manifold $M_n$ into $Z$, subject to the equivalence relation that the pair $(M_n,f)$ is corbordant to zero if and only if there is a spin $(n+1)$-manifold $B_{n+1}$ that bounds $M_n$ and an extension $\tilde{f}:B_{n+1}\rightarrow Z$ of the map $f$. The group structure is simply given by the disjoint union of manifolds.

In the case at hand, we have a spin six-manifold $S$ together with a dimension two cohomology class $a$ on $S$. The class $a$ may be defined by a map $f:S \rightarrow K(\mathbb{Z},2)$ from $S$ to the Eilenberg-MacLane space $K(\mathbb{Z},2)$ that classifies integral two-dimenional cohomology. Equivalently, such classes $a$ are in one-to-one correspondence with complex line bundles over $S$. The classifying space is ${\mcp}^{\infty} = K(\mathbb{Z},2)$. Thus, equivalently,

\be
(S,a) = 0 \in \Omega_6^{\mathrm{spin}}(\mcp^{\infty})\label{bordism}\ee

One can show that $\Omega_6^{\mathrm{spin}}({\mcp}^{\infty}) = \Omega_8^{\mathrm{spin}^c} \equiv \Omega_8^{\mathrm{spin}^c}(\{\mathrm{point}\})$ is a non-trivial group \cite{stong}, and therefore that (\ref{bordism}) is a non-trivial statement.

A key notion in (co)-bordism theory is that of a characteristic number. For example, for the map $f:M \rightarrow Z$ one has generalised Pontryagin numbers

\be
\int_M f^*(z)\cdot p_{\omega}(TM)\ee

where $p_{\omega}(TM)$ denotes a monomial\footnote{The subscript $\omega$ denotes the particular monomial.} in Pontryagin numbers of $M$, and $z$ is any element of $H^*(Z;\mathbb{Z})$. The cohomology ring $H^*({\mcp}^{\infty};\mathbb{Z})$ of the classifying space $\mcp^{\infty}$ is the integral polynomial ring on the generator $x=c_1$. Pulling this back to $M$ via the map $f$ therefore gives the first Chern class $a$. For us, this is the first Chern class of the M-theory circle bundle. In particular, we have a homomorphism $h:\Omega_6^{\mathrm{spin}}(\mcp^{\infty}) \rightarrow \mathbb{Z}$ given by

\be
h((S,a)) = \int_S a \wedge p_1(TS)\ee

If $(S,a)$ is zero as a spin bordism class, the characteristic number $h((S,a))$ must vanish. Using (\ref{oned6again}) we therefore have that

\be
h((S,a)) = \int_S a \wedge p_1(TS) = \int_B p_1(TS)\label{h}\ee

should vanish, and therefore (\ref{charge}) should vanish. Implicit in our calculation here is the fact that $p_1(TS)$ is a pull-back from $B$ (it is also the restriction of $p_1(TY)$ to $S$).

Notice that $h$ is not the only bordism invariant of $(S,a)$ - it just happens to be the invariant of interest. For example, the characteristic number

\be
\int_S a^3\ee

should also vanish in the current situation. One can prove that this is indeed the case using the results of section 3.1. The details are left as an exercise for the interested reader.

%%%%%%%%%%%%%%%%%%%%%%%%%%%%%%%%%%%%%%%%%%%%%%%%%%%%%%%%%%%%%%%%%%%%%%%%%%

\medskip

\centerline{\bf Acknowledgments}
\noindent 

It is a pleasure to thank Sergei Gukov for useful comments and discussions.

\end{document}